%% file: main-arxiv.tex
\documentclass[acmsigecom]{acmtrans2m}

\let\cite\relax 
\let\bibhang\relax

\usepackage[authoryear]{natbib} 
\usepackage{amsmath,amssymb}
\newtheorem{example}{Example}
\usepackage{xspace}
\usepackage{nicefrac}
\usepackage{booktabs}
\usepackage{graphicx}
\usepackage{tikz}
\usepackage{caption}
\usepackage{subcaption}
\usepackage{bbm}
\usepackage{pgfplots}
\pgfplotsset{compat=1.18} 
\usepgfplotslibrary{patchplots}

\renewcommand{\paragraph}[1]{\par\medskip\noindent{\textit{#1}}\xspace}

\newtheorem{theorem}{Theorem}
\newtheorem{proposition}{Proposition}

\newcommand{\rew}[0]{r\xspace}
\newcommand{\Rew}[0]{R\xspace}
\newcommand{\cost}[0]{c\xspace}
\newcommand{\costs}[0]{\textbf{c}\xspace}
\newcommand{\prob}[0]{q\xspace}
\newcommand{\Prob}[0]{\mathbf{q}\xspace}
\newcommand{\Wel}[0]{W\xspace}
\newcommand{\pay}[0]{t\xspace}
\newcommand{\Pay}[0]{T\xspace}
\newcommand{\istar}[0]{i^\star\xspace}
\newcommand{\con}[0]{\mathbf{t}\xspace}
\newcommand{\linconvec}[0]{\boldsymbol{\alpha}\xspace}

\newcommand{\indicator}[1]{\mathbbm{1}\left[{#1}\right]}
\newcommand{\reals}{\mathbb{R}}
\newcommand{\agents}{N}
\newcommand{\numrounds}{S}
\newcommand{\round}{s}

\DeclareMathOperator*{\argmax}{arg\,max}

\usepackage[suppress]{color-edits} 
\addauthor{PD}{red}

\addauthor{MF}{blue}

\addauthor{ITC}{magenta}

\newcommand{\itc}[1]{{\ITCedit{#1}}}
\addauthor{TODO}{brown}

\title[An Introduction to Contract Theory for Computer Scientists]{An Introduction to Contract Theory for Computer Scientists}

\author[PAUL D\"UTTING and MICHAL FELDMAN and INBAL TALGAM-COHEN]{ PAUL D\"UTTING\\ Google Research, Switzerland \and MICHAL FELDMAN \\ Tel Aviv University, Israel \and INBAL TALGAM-COHEN \\ Tel Aviv University, Israel}

\begin{abstract} 
This survey introduces contract design to computer scientists: using outcome-based payments to incentivize effort. We discuss applications that motivate a computational approach to contract design, ranging from online labor markets, to incentive-aware algorithmic classification, to emerging ecosystems of AI agents. We present the classic principal-agent model and introduce basic concepts. We show how optimal contracts are computable via linear programming, and why they can be opaque and brittle. Motivated by this we turn to simpler contract formats, specifically linear contracts. We discuss approximation guarantees, and robust (max–min) optimality of such contracts. We then move beyond the single-agent setting to combinatorial contract settings with multiple agents, mapping the algorithmic landscape of tractability and approximation. Finally, we discuss partial information settings through the lens of learning algorithms, and outline open directions. The goal is a concise, accessible entry point that connects economic foundations with algorithmic tools and highlights opportunities for new theory and applications. 
\end{abstract}
            
            

            
\begin{document}
            
\begin{bottomstuff} 
Appeared in ACM SIGecom Exchanges~\citep{DuettingFT26}.\\
Authors' addresses: \texttt{duetting@google.com}, \texttt{mfeldman@tau.ac.il},
\texttt{inbaltalgam@gmail.com}.
\end{bottomstuff}
            
\maketitle

\section{Introduction} 

The field of Algorithmic Game Theory (AGT) addresses the challenges and opportunities that arise when algorithms interact with real-life, strategic players. 
One of the founding papers of the field, by \cite{NisanR01}, was motivated by the rising prominence of the internet circa 2000 ---``a distributed setting where the participants cannot be assumed to follow the algorithm but rather their own self-interest''.
Nisan and Ronen suggested a framework for studying algorithms in such settings, based on \emph{mechanism design} from microeconomics. This led to a vast body of research on algorithmic mechanism design, with a focus on payment schemes that incentivize strategic players to \emph{report truthfully} to algorithms. 

However, in many instances, strategic reaction to algorithms is unrelated to truthfulness and mechanism design. In this survey we introduce \emph{algorithmic contract design}---a new frontier for AGT. Contract design is a renowned subfield of microeconomics.  
It is similar to mechanism design in its theme of incentivizing strategic agents, but the incentives are not to report truthfully, but rather to exert effort earnestly. A contract is essentially a performance-based payment scheme, which ties an agent's compensation to that agent's performance and created value.  

A premier application of contracts is in online markets of services, which, similar to markets of goods in the first two decades of the internet, have recently seen a massive growth both in importance and scale---creating a unique opportunity for computational approaches. 

\begin{example}[(Influencer marketing)]
    \label{ex:intro1}
    Consider a brand that wants to pay a social media influencer (say a famous TikToker) to post sponsored/branded content. How should the contract be structured? Should it be a flat payment? Or should the payment be tied to how successful the post is? Should it be tied to the number of views, or the number of conversions? 
\end{example} 
 
A second, perhaps less obvious applications of contract theory, is as a framework for understanding how humans react to algorithms such as a classification algorithm (even absent payments). This connection plays a role, for example, in the design of massive open online courses (MOOCs).

\begin{example}[(Strategic reaction to a Classifier)]
    \label{ex:intro2}
    Consider a classification algorithm, classifying human subjects according to some features they exhibit. This could be a prediction algorithm that assigns credit scores, or an exam at a university course or MOOC. The subjects of classification may invest strategic effort into developing more favorable features. In other words, an algorithm cannot measure without changing what it measures.  
    Contract design helps design classifiers that anticipate this strategic effect, and can help route the efforts to modify features into productive avenues. 
\end{example}

We are now on the cusp of a new revolution, where AI agents are becoming increasingly capable of autonomously carrying out more and more tasks. The classic, pre-set algorithms of Nisan and Ronen are being replaced with opaque GenAI models, operating autonomously on behalf of their users and interacting with other AI agents. 
This growing ecosystem of AI agents and LLM providers depends crucially on effective delegation and cooperation, as enabled by contract design. 

\begin{example}[(Paying an AI agent)]
    \label{ex:intro3}
    Consider a marketing GenAI agent, creating content for a campaign on behalf of a brand. What is the ideal payment scheme? Current LLM monetization by providers like OpenAI or Anthropic is usually per-token or subscription-based. In comparison, a performance-based contract would enable the utilization of key performance indicators like profit or audience engagement, thus leading to aligned incentives and elevated performance.
\end{example}

The examples demonstrate some of the fundamental computational questions that arise when considering contracts: Perhaps the simplest possible performance-based payment scheme is a fixed fraction of the profit, known as a \emph{linear} contract; how far from optimal is such a scheme, relative to a more complex one that relies on a combination of multiple indicators? 
What if instead of a single agent, 
there is a team of agents; how should they split the profit? 
What if certain aspects that are relevant to the design of contracts are not fully known, and have to be learned?
These and other challenges require a combination of algorithmic and learning techniques with economic theory, and form the new research area of algorithmic contract design. 

This survey aims to provide an accessible entry point to this new area for computer scientists. 
A more comprehensive survey by \cite{DuttingFT24} covers additional topics within this new frontier and gives additional details on the topics surveyed here. A separate survey by \cite{Feldman25}
covers in greater depth recent advances in multi-agent and multi-action  combinatorial settings.

\paragraph{Organization.} 
We introduce the canonical contract model in Section~\ref{sec:model}. In Section~\ref{sec:opt-lp} we explore optimal contracts, and discuss simple, linear contracts in Section~\ref{sec:linear}.  Section~\ref{sec:combinatorial} explores combinatorial, multi-agent contracts. 
In Section~\ref{sec:learning}, we take a learning perspective. A discussion of what's ahead appears in Section~\ref{sec:outlook}.

\section{Model}
\label{sec:model}
\input{model-short}

\section{Optimal Contracts} 
\label{sec:opt-lp}

\input{optimal-short}

\section{Linear Contracts} 
\label{sec:linear}

A contract is \emph{linear} if it specifies a fixed fraction $\alpha \in [0,1]$ of the rewards to be transferred to the agent by the principal. The parameter $\alpha$ determines all the payments: $\pay_j = \alpha\rew_j$ for every outcome $j \in [m]$. The agent's expected utility from the $i$th action is then $\alpha R_i-c_i$. So as $\alpha$ becomes closer to~$1$, the agent puts more weight on the expected reward $R_i$ relative to the cost $c_i$ when choosing an action. At the extreme $\alpha=1$, the agent chooses the welfare-maximizing action. Linear contracts have several nice properties like simplicity, interpretable payments (e.g., paying per conversion in Example~\ref{ex:intro1}), and monotonicity. Due to their widespread use, linear contracts were extensively studied in the economic literature. The computational perspective provides new insights, underscoring the potential of an algorithmic theory of contracts. We describe here two such insights: the approximation guarantees of linear contracts, and their robust optimality to uncertainty about distributions. 

\begin{figure}
    \centering
    \begin{subfigure}[t]{0.4\textwidth}
    \centering
    \input{agent-utility.tex}
    \caption{The agent's perspective}
    \label{fig:upper-envelope-agent}%
    \end{subfigure}
    \hspace{1.5cm}
    \begin{subfigure}[t]{0.4\textwidth}
    \centering
    \input{principal-utility.tex}
    \caption{The principal's perspective}
    \label{fig:upper-envelope-principal}%
    \end{subfigure}
    \caption{The agent's expected utility as a function of the linear contract's parameter $\alpha$ {\bf (left)}, and the principal's expected  utility as a function of $\alpha$ {(\bf right)}, for the principal-agent setting in Example~\ref{ex:running}. 
    }
    \label{fig:upper-envelope}
\end{figure}

\paragraph{A Geometric Perspective.} 
An important tool in the analysis of linear contracts, referred to as the geometric approach~\citep{DuttingRT19}, utilizes the fact that the agent's utility $\alpha \Rew_i - c_i$ for each action $i$ is a linear function in $\alpha$. Specifically, the agent's best response to each contract $\alpha$ is given by the action $i$ that results in the highest utility for the agent at that given $\alpha$. In other words, by tracing the \emph{upper envelope} of the set of functions $\{\alpha \Rew_i - c_i\}_{i \in [n]}$ we can determine the agent's best response for each $\alpha \in [0,1]$. Figure~\ref{fig:upper-envelope-agent} illustrates this for Example~\ref{ex:running}. From this perspective, it becomes clear that the unit interval of possible linear contracts will be divided into contiguous intervals where the agent's choice of action is constant. Also, if we rename the actions as they appear on the upper envelope (from left to right), then actions are sorted by non-decreasing welfare, reward, and cost. 

Switching the roles, when we consider the principal's utility as a function of the contract $\alpha \in [0,1]$, we arrive at a graph like the one in Figure~\ref{fig:upper-envelope-principal} (which is for the same example). We see that the principal's utility is a piece-wise linear function. Each segment  corresponds to a different best-response of the agent, and within the segment corresponding to action $i$ the principal's utility decreases at rate $-\Rew_i$. Finally, under the standard tie-breaking rule, at the left-most point of a segment the agent will always choose the action with higher welfare (higher index, after re-indexing). Thus, the principal's utility is continuous when going to the right, while it might ``jump'' when going to the left.  In particular, the best way to incentivize action $i$ is by choosing the $\alpha_i$ that corresponds to the left-most endpoint of action $i$'s interval. By checking the principal's utility at all \emph{critical values} of $\alpha$, i.e., all endpoints where the agent switches from one action to another, we can determine an optimal linear contract.

\medskip
\textsc{Example~\ref{ex:running}, Continued.} \emph{Example~\ref{ex:running} demonstrates the suboptimality of linear contracts, which can obtain no more than $3$ as expected utility for the principal (see Figure~\ref{fig:upper-envelope-principal}), in comparison to $3.5$ by the optimal nonlinear contract.}  

\paragraph{Approximation Guarantees.} 
A key insight of the algorithmic study of contracts is the following set of tight (multiplicative) approximation guarantees,
showing that linear contracts provide a good approximation except in cases where there is simultaneously a large number of actions, a large spread in expected rewards, and a large spread in costs. 

\begin{theorem}[(\citet*{DuttingRT19})] \label{thm:wc-approx} 
Let $\rho$ denote the worst-case ratio between the principal's expected utility under the optimal contract, and the principal's expected utility under the optimal \emph{linear} contract. Then among all principal-agent settings with $n$ actions, $\rho = n$; among all settings with a ratio of $H$ between the highest and lowest expected reward, $\rho = \Theta(\log H)$; and among all settings with a ratio of $C$ between the highest and lowest cost, $\rho = \Theta(\log C)$. \end{theorem}

The upper bounds in Theorem~\ref{thm:wc-approx} in fact apply against the stronger benchmark of optimal welfare, while the lower bounds apply even under the regularity assumption of MLRP.  

The direction that $\rho \le n$ follows from the characterization of optimal linear contracts above.
The tightness direction ($\rho \ge n$) is established for a particular family of contract settings with $n$ actions known as the \emph{equal revenue} setting, parameterized by a small $\epsilon>0$ . This setting has exponentially increasing expected rewards among the actions: $\Rew_i = 1/\epsilon^{i-1}$; as well as exponentially increasing costs: $\cost_i = \Rew_i - i + \epsilon(i-1)$; but the welfare of the actions increases linearly: $\Wel_i = i - \epsilon(i-1)$.
The key observation is that any linear contract extracts utility of $1$ for the principal, while the setting can be set up such that a general contract can extract $W_n\approx n$. The other bounds follow by similar constructions.

\paragraph{Robust Optimality.}
A different approach, spearheaded in the context of contract design by \cite{Carroll15}, is the robust (or max-min) optimality approach. This approach assumes that certain aspects of the contracting problem are 
\emph{known} to the principal, while other aspects are \emph{unknown}.
The principal then aims to design a contract $\con$ that achieves the best\itc{-}possible worst-case utility over all instances $\mathcal{I}$ compatible with what is known. 
That is, $\con$ maximizes $U_P(\mathcal{I}\mid \con) = \inf_{I\in \mathcal{I}} U_P(I \mid \con)$, where $I\in \mathcal{I}$ is the worst-case compatible instance. 
It turns out that under a broad range of assumptions, it is a \emph{linear} contract that achieves optimal such worst-case guarantees. 

In \citeauthor{Carroll15}'s work, the aspect known to the principal is a subset of actions $A_0$ available to the agent, while the actual set of actions $A$ available to the agent can be any superset of $A_0$.

\begin{theorem}[(\cite{Carroll15})]
Linear contracts are max-min optimal when the principal knows a subset of actions $A_0$ available to the agent, and $\mathcal{I}$ includes instances in which the full action set is any superset $A$, i.e.,~$A_0\subseteq A$. 
\end{theorem}

An alternative formulation of the robustness of linear contracts was given by \citet{DuttingRT19}, in a natural model of moment information that is easy to interpret. 

\begin{theorem}[(\cite*{DuttingRT19})]
Linear contracts are max-min optimal when the principal knows the costs $c_1, \ldots, c_n$, the rewards $r_1,\dots,r_m$, and the expected rewards $\Rew_1,\dots,\Rew_n$ available to the agent, and $\mathcal{I}$ includes all instances in which the distributions over rewards induced by the actions are compatible with the expected rewards.
\end{theorem}

In both cases, the proof approach is based on showing that whatever non-linear contract $\con$ is considered as a candidate, an instance $I\in\mathcal{I}$ and a linear contract $\con'$ exist such that $U_P(I \mid \con) \le U_P(\mathcal{I} \mid \con')$. Combined with $U_P(\mathcal{I} \mid \con) \le U_P(I \mid \con)$, we get that $U_P(\mathcal{I} \mid \con) \le U_P(\mathcal{I} \mid \con')$, as desired.

\section{Combinatorial Contracts} 
\label{sec:combinatorial}

In the basic contracting model (Section~\ref{sec:model}), a single principal interacts with a single agent, who chooses one of $n$ actions leading to one of $m$ outcomes, and the revenue-optimal contract can be computed efficiently via LPs (see Section~\ref{sec:opt-lp}).

Real-world contract settings, however, are rarely 
this simple. A brand may engage multiple influencers (agents), each promoting multiple brands (principals) across various platforms (actions), evaluated by diverse metrics (outcomes).
Such generalizations introduce new computational challenges, which the growing body of work on algorithmic contract theory aims to address. 
This body of work provides both efficient algorithms and impossibility results for settings with multiple agents \citep{BabaioffFNW12,DuettingEFK23},
combinatorial actions
\citep{DuettingEFK21,DuettingFG23,VuongDPP23,DuettingFGR24}, 
combinatorial outcomes \citep{DuttingRT21}, 
multiple principals \citep{AlonLST23}, and combinations thereof~\citep{DuettingEFK24}. 
As is typical at the interface of economics and computation, the computational perspective reveals key structural insights.
Below we present a single paradigmatic example to illustrate the kinds of challenges that arise, and approaches to address them.

\paragraph{Paradigmatic Example: Multiple Agents.}
A natural extension of the contracting problem arises when a principal seeks to incentivize a team of agents. 
The classic reference here is \citet*{Holmstrom82} on moral hazard in teams. The effectiveness of a team depends heavily on its composition, leading to the algorithmic challenge of designing (near-)optimal contracts for subsets of agents. This is already nontrivial in the basic setting where agents can either exert effort or not (the binary action setting), reducing to the question of which agents to contract with. 

This line of work was initiated by \citet{BabaioffFNW12}
under the name {\em combinatorial agency}. We present here the more general model of \cite{DuettingEFK23}. There is a set $\agents$ of $n$ agents. 
Each agent $i \in \agents$ incurs a cost 
$\cost_i \in \reals_{\geq 0}$ to exert effort, and the project either succeeds (reward 1) or fails (reward 0). A set function 
$f:2^{\agents} \rightarrow [0,1]$ specifies the success probability as a function of the subset of agents exerting effort.
As in the single agent case, in this binary-outcome model one can focus on linear contracts: a vector 
$\linconvec=(\alpha_1, \ldots, \alpha_n)$, where $\alpha_i$ is the share of the reward agent $i$ receives if the project succeeds. Given a contract $\linconvec$
and a set 
$S \subseteq \agents$ of agents who exert effort, the principal’s utility is $U_P(S \mid \linconvec) := (1-\sum_{i \in \agents} \alpha_i) f(S)$, and agent $i$'s utility is $U_{i}(S \mid \linconvec) :=  \alpha_i f(S) - \indicator{i \in S} \cdot c_i$, where $\indicator{i \in S} = 1$ if $i \in S$ and $\indicator{i \in S} = 0$ otherwise.

Each contract thus defines a game among the agents.
We focus on pure Nash equilibria, namely action profiles where no agent has an incentive to deviate. A set $S$ is an equilibrium under a contract $\linconvec$ if: (i) for every $i \in S$, $\alpha_i f(S) - c_i \geq \alpha_i f(S \setminus \{i\})$ (no incentive to deviate to shirking), and
(ii) for every $i \not\in S$, $\alpha_i f(S) \geq \alpha_i f(S \cup \{i\}) - c_i$ (no incentive to deviate to exerting effort).
It is not too difficult to show that every contract $\linconvec$ admits an equilibrium~\citep{VuongDPP23,DuettingEFK24}. 
The problem is to find the contract $\linconvec$ that maximizes the principal's utility, in the best equilibrium of $\linconvec$.

The optimal choice of $\linconvec$ for a given set $S$ is $\alpha_i = \frac{c_i}{f(i \mid S \setminus \{i\})}$ for $i \in S$, and $\alpha_i = 0$ for $i \notin S$.
That is, the payment to an agent in $S$ is proportional to his cost and inverse proportional to his marginal contribution to the success probability, reflecting the fact that an agent whose work matters less requires high compensation to avoid free riding on others' work.
Thus, finding the optimal contract reduces to identifying the set $S^\star$ that maximizes the function
\[
g : 2^{\agents} \to \mathbb{R} \cup \{-\infty\}, \qquad \text{where} \quad 
g(S) := \left(1 - \sum_{i \in S} \frac{c_i}{f(i \mid S \setminus \{i\})} \right) f(S).
\]

\paragraph{Approximation Results.}
\citet{DuettingEFK23} study the problem of computing near-optimal contracts, assuming the success function $f$ belongs to the hierarchy of complement-free set functions (e.g., additive, submodular, XOS, subadditive), capturing situations where an agent contribution diminishes as the team grows. They show that even for additive $f$, the problem is \textsf{NP}-hard,
but admits an FPTAS. 
For more general set functions, they establish approximation results, given access to common primitives: A value query is given a set $S$ and returns $f(S)$. A demand query gets a price vector $p_1,\ldots,p_m$ and returns a set maximizing $f(S)-\sum_{j \in S}p_j$. 

\begin{theorem}
[(\citet*{DuettingEFK23})]
\label{thm:multi-agents}
For submodular (resp., XOS) success functions $f$, an $O(1)$-approximation to the optimal contract can be computed using poly-many value queries (resp., value and demand queries). 
\end{theorem}

The proof of Theorem~\ref{thm:multi-agents} reveals a (perhaps surprising) connection 
to pricing. 
Namely, it can be shown that using a demand query
with carefully chosen prices allows one to identify a ``good'' set of agents---one whose success value $f(S)$ is at least half that of the optimal set.
However, to ensure a high $g(S)$ value, the payments must remain low; in particular, agents' marginal contributions to $f$ must be sufficiently large. 
The crux is therefore a scaling lemma, which shows that it is always possible to remove agents from the set $S$, 
while also preserving a sufficiently large fraction of the original marginal contributions. For submodular functions, the latter part is straightforward, since the marginal contributions of the remaining agents can only go up. For XOS functions, however, the marginal contributions of the  agents that remain may decrease, so a more careful argument is required to establish such a scaling property.

This proof pattern yields an efficient algorithm with value and demand queries. For submodular functions, the argument can be further refined, by replacing demand queries with an appropriately defined relaxation of demand queries that can be computed with poly-many value queries.

\paragraph{Impossibility Results.}
Follow-up work by \citet{EzraFS24} shows that for submodular functions, there exists a constant $c > 1$ such that no polynomial-time algorithm with value oracle access can approximate the optimal contract to within a factor better than $c$, assuming \textsf{P}$\neq$\textsf{NP}.
More recently, \citet{DuettingEFK24} showed that even with both value and demand oracle access to the submodular function, there exists a constant $c > 1$, such that any algorithm that uses a sub-exponential number of queries returns a $c$-approximation with probability exponentially-small in~$n$. 
In addition, for XOS functions, \citet{EzraFS24} show that no
algorithm that makes poly-many value queries can approximate the optimal contract (with high probability) to within a factor 
$\Omega(n^{1/6})$.

\section{Learning Contracts}
\label{sec:learning}

Another natural direction for an algorithmic study of contracts pioneered in \cite{HoSV16}, is to study the contract design problem from a learning perspective.
Here we focus on the online learning problem with bandit feedback studied in \cite{HoSV16}, and discuss the state-of-the-art results of \citet{ZhuEtAl22}. Other variants of the problem have been studied, including under additional restrictions on the model (small action spaces, distributions satisfying additional conditions, etc.)~\citep{BacchiocchiC0024,ChenCDH24}, and under different feedback models~\citep{DuettingGSW23,ChenCDH24,sample-complexity}. Finally, \cite{WangEtAl2023} propose a deep neural network architecture for designing near-optimal contracts from samples. 

\input{learning-short}

\section{Looking Ahead}
\label{sec:outlook}

Attesting to the rapid growth of the new research area of algorithmic contract design, in this brief survey we have not been able to touch upon all lines of research. For example, there has been much interest in \emph{typed} contract design: Bayesian environments where agents have private type information, which call for a unified treatment of mechanism design and contract design \citep{GuruganeshSW21,AlonDT21,AlonDLT23,CastiglioniCL+25,CastiglioniM022}.

While substantial progress has been made in developing the foundations of algorithmic contract design, several directions remain only partially explored. These  additional directions include 
inspection and monitoring of actions and outcomes \citep{EzraLR24}, the interplay with information revelation \citep{BabichenkoEtAl2024,CastiglioniC25}, ambiguous contract design \citep{DuettingFP23,DuettingFR25}, sequential decision making \citep{ezra2024sequentialcontracts}, 
and classification \citep{KleinbergR19,AlonDPTT20} (as in Example~\ref{ex:intro2}).
It would also be interesting to further explore the connection to closely related approaches such as scoring rules \citep{HartlineSLW22} 
and delegation \citep{KleinbergK18}.

Finally, circling back to Example~\ref{ex:intro3}, the rise of an ``agentic era," where autonomous AI agents act as both principals and agents, raises new challenges only beginning to be addressed~\citep{SaigET24}. Vast and dynamic action spaces, shifting assumptions about rationality and commitment, and the computational capabilities of AI agents open fresh avenues for contract design. As in algorithmic mechanism design, we expect the computational lens to reveal new structures, tractability frontiers, and insights, shaping the next phase of contract theory. 

\section*{Acknowledgments}
This survey received funding from the European Research Council (ERC) under the European Union's Horizon 2020 research and innovation program (grant agreement No.~101077862) and the Horizon Europe program for research and innovation (grant agreement No.~101170373), by the Israel Science Foundation (grant No.~3331/24), by the Israel Science Foundation Breakthrough Program (grant No.~2600/24), by the NSF-BSF (grant No.~2020788 and grant No.~2021680), by a Google Research Scholar Award, and by an Amazon Research Award. 

\bibliographystyle{acmtrans}
\bibliography{bib-short}

\begin{received}
\end{received}
\end{document}

%% file: model-short.tex
In this section, we introduce the hidden-action principal agent model \citep{Holmstrom79,GrossmanH83}. 
Our coverage follows \citep{DuttingRT19}. 

\paragraph{Setting.} 
There is a principal (she) and an agent (he). 
The agent can take one of $n$ actions. 
Each action $i \in \mathcal{A} = [n]$ has a cost $c_i \geq 0$. Actions are sorted by non-increasing cost. The cost of an action is borne by the agent. 
We model the fact that the agent can opt out of the contract 
by assuming that one of the actions has zero cost. 
Each action stochastically leads to one of $m$ outcomes. We write $\prob_{ij}$ for the probability of outcome $j$ under action $i$, and use $\Prob_i$ to denote the probability distribution over outcomes induced by action $i$. 
The agent's choice of action is \emph{hidden} from the principal, who can  only observe the stochastic outcome of the action. This reflects that 
the principal is unable or unwilling to keep track of the agent's action.
Each outcome $j$ is associated with a reward $r_j \geq 0$, enjoyed by the principal. Rewards are sorted in non-increasing order. 
We use
$\Rew_i := \mathbb{E}_{j\sim \Prob_i}[\rew_j] =\sum_{j\in [m]} \prob_{ij} \rew_j$
to denote the expected reward of action $i$. 

\begin{example}[(Running Example)]
\label{ex:running}
Consider a principal-agent setting with three actions $i\in[3]$ and three outcomes~$j\in[3]$:
\smallskip
\begin{center}
\begin{tabular}{|l|lll||c|}
\toprule
reward: & $r_1 = 0$ & $r_2 = 1$ & $r_3 = 7$ & \text{cost}\\
\midrule 
action $1$: & $\prob_{11} = 1$ & $\prob_{12} = 0$ & $\prob_{13} =0$ & $c_1 = 0$\\
action $2$: & $\prob_{21} = 0$ & $\prob_{22} = \nicefrac{1}{2}$ & $\prob_{23} = \nicefrac{1}{2}$ & $\cost_
2 = 1$ \\ 
action $3$: & $\prob_{31} = 0$ & $\prob_{32} = \nicefrac{1}{6}$ & $\prob_{33} = \nicefrac{5}{6}$ & $\cost_3 = 2$\\
\bottomrule
\end{tabular}
\end{center}
\medskip
If the agent takes action $2$, then he bears a cost of $1$. The outcome is $j = 2$ or $j = 3$ with equal probability. Note that if the principal observes outcome $2$, she cannot be certain this is because the agent chose action $2$, but action $2$ is more likely than action $3$. 
\end{example}

It is possible to impose additional structure on the distributions, like (first-order) stochastic dominance, which reflects that more rewarding outcomes become more likely as the agent exerts more costly effort. 
Other common assumptions include the stronger
\emph{Monotone Likelihood-Ratio Property} (MLRP), or the \emph{Concavity of Distribution Function Property} (CDFP).

\paragraph{Contract.} 
The principal enjoys the rewards while the agent bears the cost; these misaligned preferences create an incentive problem called \emph{moral hazard}, which the principal seeks to address by defining a contract.  
A contract is a payment rule $\con$ that consists of $m$~non-negative payments or \emph{transfers} $(\pay_1,...,\pay_m)$, one for each outcome.%
\footnote{The requirement that payments should be non-negative is referred to as \emph{limited liability}. It reflects the asymmetric roles of the principal and the agent in contractual relations, 
and also rules out trivial but unrealistic solutions.
}   
The transfers are associated with outcomes rather than actions since the actions are hidden from the principal.
For action $i\in[n]$ let
\begin{equation}
\Pay_i := \mathbb{E}_{j \sim \Prob_i}[\pay_j] =\sum_{j\in [m]} \prob_{ij} \pay_j\label{eq:expected-transfer}
\end{equation}
denote the expected payment from principal to agent for taking action $i$.

Both the principal and the agent are assumed to be \emph{risk neutral}. For a fixed contract $\con$ and choice of action $i$, the agent's expected utility and principal's expected utility (a.k.a.~\emph{revenue}) are given by
\[
U_A(i \mid \con) := \Pay_i - c_i \quad\text{and}\quad U_P(i \mid \con) := \Rew_i - \Pay_i.
\]

Notice that the sum of the players' expected utilities is always equal to the expected welfare $\Wel_i := \Rew_i-\cost_i$ of the action $i$ chosen by the agent. 
The contract thus influences the agent's choice of the welfare ``pie'' (through his choice of action), in addition to determining how this pie is divided between the principal and the agent.

\medskip
\textsc{Example~\ref{ex:running}, Continued.} \emph{Consider the contract $\con = (0,1,3)$. 
The expected payment for action~$1$ under this contract is $\Pay_1 = 0$, for action $2$ it is $\Pay_2 = 2$, and for action $3$ it is 
$\Pay_3 = \nicefrac{8}{3}$. The agent's expected utility is thus maximized by action $2$, which yields an expected utility of $U_A(2\mid\con) = 1$, compared to an expected utility of $U_A(1 \mid \con) = 0$ for action $1$ and an expected utility of $U_A(3 \mid \con) = \nicefrac{2}{3}$ for action $3$. The principal's utility is then $U_p(2 \mid \con) = 2.$} 

\begin{figure}[t]
\vspace*{-10pt}
\centering
\include{timing-contracts-classic.tex}
\vspace*{-28pt}
\caption{Timeline.}\label{fig:timeline}
\end{figure}

\paragraph{Incentives.}
The contract design problem defines a game between the principal and the agent, in which the principal moves first and defines a contract $\con$ and the agent reacts with a utility-maximizing action $i^\star(\con)$ (see Figure~\ref{fig:timeline}).
Formally, for a fixed contract $\con$  
let $\mathcal{A}^\star(\con) := \argmax_{i\in [n]} U_A(i \mid \con) \subseteq [n]$ denote the set of actions that maximize the agent's expected utility. Using this notation, the agent chooses an action
\begin{align}
\istar\in \mathcal{A}^\star(\con) = \argmax_{i\in [n]}\; U_A(i \mid \con).
\label{eq:istar}
\end{align}
Any such choice $\istar$ is \emph{incentive compatible} (IC) for the agent, because it is preferred over any other action. 
Note that it is also \emph{individually rational} (IR) for the agent, meaning that the agent is guaranteed non-negative utility. This is because payments are non-negative, and we assumed that the agent has a zero-cost action.

Importantly, the principal's utility may be different for different actions $i^\star \in \mathcal{A}^\star(\con)$. 
It is thus important to specify how the agent breaks ties. 
By default, and as is standard in the contracts literature, 
we assume that the agent breaks ties in favor of the principal.\footnote{
This tie-breaking rule is justified by the fact that a small perturbation would make the agent strictly prefer that action (see, e.g., \cite{Carroll15,DuttingRT19} for additional discussion).} 
We refer to this tie-breaking rule as the \emph{canonical} tie-breaking rule.

%% file: timing-contracts-classic.tex
\scalebox{0.95}{
\begin{tikzpicture}
\draw[blue, thick,->] (0,0) -- (11,0) node[right]{\textcolor{blue}{Time}};
\draw[black,thick,->] (0.5,-0.5) -- (0.5,0);
\node at (0.5,-1) {\small{Principal \textcolor{black}{offers} agent}};
\node at (0.5,-1.4) {\small{a contract}};
\node at (0.5,-1.9) {\small{(parties have}};
\node at (0.5,-2.3) {\small{symmetric info)}};
\node at (0.5,1) {\includegraphics[scale=0.4]{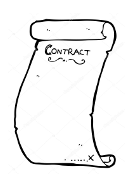}};
\draw[black,thick,->] (3,-0.5) -- (3,0);
\node at (3,-1) {\small{Agent}};
\node at (3,-1.4) {\small{\textcolor{black}{accepts}}};
\node at (3,-1.9) {\small{(or refuses)}};
\node at (3,1) {\includegraphics[scale=0.4]{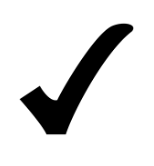}};
\draw[black,thick,->] (5.5,-0.5) -- (5.5,0);
\node at (5.5,-1) {\small{Agent takes}};
\node at (5.5,-1.4) {\small{costly,}};
\node at (5.5,-1.8) {\small{\textcolor{black}{hidden}}};
\node at (5.5,-2.2) {\small{\textcolor{black}{action}}};
\node at (5.5,1) {\includegraphics[scale=0.4]{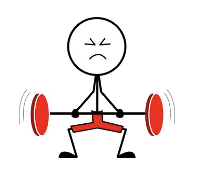}};
\draw[black,thick,->] (8,-0.5) -- (8,0);
\node at (8,-1) {\small{Action's}};
\node at (8,-1.4) {\small{outcome}};
\node at (8,-1.8) {\small{\textcolor{black}{rewards} the}};
\node at (8,-2.2) {\small{principal}};
\node at (8,1) {\includegraphics[scale=0.4]{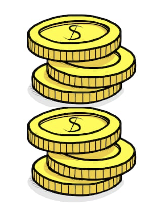}};
\draw[black,thick,->] (10.5,-0.5) -- (10.5,0);
\node at (10.5,-1) {\small{Principal}};
\node at (10.5,-1.4) {\small{\textcolor{black}{pays} agent}};
\node at (10.5,-1.8) {\small{according}};
\node at (10.5,-2.2) {\small{to contract}};
\node at (10.5,1) {\includegraphics[scale=0.4]{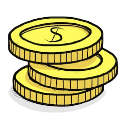}};
\end{tikzpicture}
} 

%% file: optimal-short.tex
The principal's design problem is to select a contract $\con$ that maximizes her expected utility (i.e., revenue from the contract), given that the agent chooses an action $\istar \in \mathcal{A}^\star(\con)$ that maximizes his own expected utility. This contract is referred to as \emph{optimal}.
The next result is an algorithmic reformulation of a characterization of optimal contracts by \citep{GrossmanH83}, showing that the optimal contract can be computed (in polyomial time) by solving 
$n$ \emph{linear programs} (LPs), one per action. 

\begin{proposition}[(\cite{GrossmanH83})]\label{prop:LP}
    An optimal contract can be found by solving $n$ linear programs, one per action. 
\end{proposition}

To outline the LP approach, we call an action 
$i \in [n]$ \emph{implementable} (up to tie-breaking) if there exists a contract 
$\con$ such that
$U_A(i \mid \con)  
\geq U_A(i' \mid \con)$ 
for every $i' \neq i$.
The approach is to construct an LP for each action 
$i \in [n]$ that determines whether the action is implementable 
and, if so, to find the minimum expected payment required to implement it. 
This LP and its dual are given in Figure~\ref{fig:primal-dual-lp}. 
An optimal contract can then be found by applying this procedure to all actions, and choosing the action that maximizes the principal's utility. The corresponding pair of action and contract can be shown to be compatible with the canonical tie breaking rule.

An important corollary of the LP formulation 
is a characterization of the actions that the principal can implement. 

\begin{proposition}[(\cite{HermalinK91})]
\label{prop:implementable}
Action $i$ is implementable if and only if there is no convex combination $\{\gamma_{i'}\}_{i' \neq i}$ of the actions other than $i$ that results in the same distribution over outcomes, i.e., $\sum_{i' \neq i} \gamma_{i'} \prob_{i'j} = \prob_{i,j}$ for all outcomes $j$, with lower weighted cost, i.e., $\sum_{i' \neq i} \gamma_{i'} \cost_{i'} < \cost_i$.
\end{proposition}

The characterization follows from LP duality: \textsf{MINPAY-LP($i$)} is infeasible (equiv., action $i$ cannot be implemented) precisely when its dual (which is always feasible) is unbounded. If
action $i$'s outcome distribution can be replicated by a lower-cost convex combination of other actions, then a scaled version of the coefficients of this combination both satisfies the dual constraints, and achieves arbitrarily-high dual objective. The opposite direction is also not hard to show, establishing the characterization.

\paragraph{Optimal Contracts in Special Cases.}
In the special cases of binary actions and binary outcomes, optimal contracts have a particularly simple structure. 
For binary outcomes, the optimal contract is linear (see Section~\ref{sec:linear}). For binary actions, the optimal contract only pays for a single outcome, the one that maximizes the so-called likelihood ratio. 
This also holds more generally for instances that\itc{,} beyond the two actions\itc{,} have a zero-cost action that leads to a zero-reward outcome with certainty, as in our running example. 

\begin{figure}[t]
\begin{center}
\begin{subfigure}[t]{0.45\textwidth}
\footnotesize
\begin{align*}
\min_{}
\quad &\sum_{j} \prob_{ij} \pay_j\\
\text{s.t.}\quad &\sum_{j} \prob_{ij} \pay_j - \cost_{i} \geq \sum_{j} \prob_{i'j} \pay_j - \cost_{i'} &&\forall i' \neq i  \\
     &t_j \geq 0 &&\forall j 
\end{align*}
\caption{\textsf{MINPAY-LP($i$)}}
\label{fig:primal-lp}
\end{subfigure}
~
~
\begin{subfigure}[t]{0.45\textwidth}
\footnotesize
\begin{align*}
\max_{}
\quad &\sum_{i' \neq i} \lambda_{i'}(\cost_i-\cost_{i'})\\
\text{s.t.}\quad &\sum_{i' \neq i} \lambda_{i'} (\prob_{ij} - \prob_{i'j}) \leq \prob_{ij} &&\forall j\\
     &\lambda_{i'} \geq 0 &&\forall i' \neq i
\end{align*}
\caption{\textsf{DUAL-MINPAY-LP($i$)}}
\label{fig:dual-lp}
\end{subfigure}
\end{center}
\vspace*{-8pt}
\caption{The \textsf{MINPAY-LP($i$)} for action $i$ \textbf{(left)} and its dual \textbf{(right)}. The primal variables are $\pay_j$ for $j \in [m]$, and the dual variables are $\lambda_{i'}$ for $i' \in [n] \setminus \{i\}$. 
}\label{fig:primal-dual-lp}
\end{figure}

\medskip
\textsc{Example~\ref{ex:running}, Continued.} \emph{The optimal contract in our running example incentivizes the agent to take action $3$, with contract $t = (0,0,3)$ which has a non-zero payment only for outcome $3$. Note that outcome $3$ is the outcome that maximizes the likelihood ratio $\prob_{3j}/\prob_{2j}$ over all outcomes $j$.}

\paragraph{Shortcomings of Optimal Contracts.}
Beyond special cases, optimal contracts are often opaque and generally do not lend themselves to a clear, intuitive interpretation.
Moreover, they are known to exhibit certain counterintuitive properties from an economic perspective. One particularly notable issue is that optimal contracts are not necessarily \emph{monotone}, meaning that in the optimal contract $\con$, a higher principal reward $\rew_j$ may entail a lower payment $\pay_j$.
This is unnatural as the agent would be paid less, even though he achieved a better outcome.

Another drawback of optimal contracts is their reliance on perfect knowledge of inputs, such as distributions $\Prob$ and costs $\costs$. This issue is exacerbated by the fact that they are sensitive to small perturbations. These shortcomings motivate some of the work covered in the following sections.

%% file: agent-utility.tex
\scalebox{0.8}{
\begin{tikzpicture}
    \useasboundingbox (0,-2) rectangle (5,3);
    \draw[thick,->] (0,-1.5) -- (0,2.5) node[above] {\footnotesize $\alpha \cdot \Rew_i - \cost_i$};
    \draw[thick,->] (0,0) node[left] {\tiny $0~$} -- (5.5,0) node[right] {\footnotesize $\alpha$};
    \draw[thick,-] (0.5,0.05) -- (0.5,-0.05); 
    \draw[thick,-] (1,0.1) -- (1,-0.1) node[below] {\tiny $0.2$}; 
    \draw[thick,-] (1.5,0.05) -- (1.5,-0.05);
    \draw[thick,-] (2,0.1) -- (2,-0.1) node[below] {\tiny $0.4$};
    \draw[thick,-] (2.5,0.05) -- (2.5,-0.05);
    \draw[thick,-] (3,0.1) -- (3,-0.1) node[below] {\tiny $0.6$};
    \draw[thick,-] (3.5,0.05) -- (3.5,-0.05);
    \draw[thick,-] (4,0.1) -- (4,-0.1) node[below] {\tiny $0.8$};
    \draw[thick,-] (4.5,0.05) -- (4.5,-0.05);
    \draw[thick,-] (5,0.1) -- (5,-0.1) node[below] {\tiny $1.0$};
    \draw[thick,-] (0.1,-1.0) -- (-0.1,-1.0) node[left] {\tiny $-2$};
    \draw[thick,-] (0.1,-0.5) -- (-0.1,-0.5) node[left] {\tiny $-1$};
    \draw[thick,-] (0.1,0.5) -- (-0.1,0.5) node[left] {\tiny $1$};
    \draw[thick,-] (0.1,1.0) -- (-0.1,1.0) node[left] {\tiny $2$};
    \draw[thick,-] (0.1,1.5) -- (-0.1,1.5) node[left] {\tiny $3$};
    \draw[thick,-] (0.1,2.0) -- (-0.1,2.0) node[left] {\tiny $4$};
    \draw[line width=2.5pt,black] (0,0) -- (1.25,0);
    \node at (0.625,-1.2) {\scriptsize \textcolor{black}{action $1$}};
    \draw[thick,blue,dashed] (1.25,-1.5) -- (1.25,2.5);
    \draw[thick,blue] (0,-0.5) -- (5,1.5) node[right] {\scriptsize{$\Wel_2 = \Rew_2 - \cost_2$}};
    \node at (-0.9,-0.5) {\textcolor{blue}{\scriptsize{$-\cost_2$}}};
    \draw[line width=2pt,blue] (1.25,0) -- (2.5,0.5);
    \node at (1.875,-1.2) {\scriptsize \textcolor{blue}{action $2$}};
    \draw[thick,red,dashed] (2.5,-1.5) -- (2.5,2.5);
    \draw[thick,red] (0,-1) -- (5,2) node[right] {\scriptsize{$\Wel_3 = \Rew_3 - \cost_3$}};
    \node at (-0.9,-1) {\textcolor{red}{\scriptsize{$-\cost_3$}}};
    \draw[line width=2pt,red] (2.5,0.5) -- (5,2);
    \node at (3.75,-1.2) {\scriptsize \textcolor{red}{action $3$}};
\end{tikzpicture}
} 

%% file: principal-utility.tex
\scalebox{0.8}{
\begin{tikzpicture}
    \useasboundingbox (0,-2) rectangle (5,3);
    \draw[thick,->] (0,-1.5) -- (0,2.5) node[above] {\footnotesize $(1-\alpha) \Rew_{i^\star(\alpha)}$};
    \draw[thick,->] (0,0) node[left] {\tiny $0~$} -- (5.5,0) node[right] {\footnotesize $\alpha$};
    \draw[thick,-] (0.5,0.05) -- (0.5,-0.05); 
    \draw[thick,-] (1,0.1) -- (1,-0.1) node[below] {\tiny $0.2$}; 
    \draw[thick,-] (1.5,0.05) -- (1.5,-0.05);
    \draw[thick,-] (2,0.1) -- (2,-0.1) node[below] {\tiny $0.4$};
    \draw[thick,-] (2.5,0.05) -- (2.5,-0.05);
    \draw[thick,-] (3,0.1) -- (3,-0.1) node[below] {\tiny $0.6$};
    \draw[thick,-] (3.5,0.05) -- (3.5,-0.05);
    \draw[thick,-] (4,0.1) -- (4,-0.1) node[below] {\tiny $0.8$};
    \draw[thick,-] (4.5,0.05) -- (4.5,-0.05);
    \draw[thick,-] (5,0.1) -- (5,-0.1) node[below] {\tiny $1.0$};
    \draw[thick,-] (0.1,-1.0) -- (-0.1,-1.0) node[left] {\tiny $-2$};
    \draw[thick,-] (0.1,-0.5) -- (-0.1,-0.5) node[left] {\tiny $-1$};
    \draw[thick,-] (0.1,0.5) -- (-0.1,0.5) node[left] {\tiny $1$};
    \draw[thick,-] (0.1,1.0) -- (-0.1,1.0) node[left] {\tiny $2$};
    \draw[thick,-] (0.1,1.5) -- (-0.1,1.5) node[left] {\tiny $3$};
    \draw[thick,-] (0.1,2.0) -- (-0.1,2.0) node[left] {\tiny $4$};
    \draw[black,line width=2pt] (0,0) -- (1.25,0);
    \node at (0.625,-1.2) {\scriptsize \textcolor{black}{action $1$}};
    \draw[thick,blue,dashed] (1.25,-1.5) -- (1.25,2.5);
    \node at (1.875,-1.2) {\scriptsize \textcolor{blue}{action $2$}};
    \draw[blue,line width=2pt] (1.25,1.5) -- (2.5,1);
    \draw[thick,red,dashed] (2.5,-1.5) -- (2.5,2.5);
    \draw[red,line width = 2pt] (2.5,1.5) -- (5,0);
    \node at (3.75,-1.2) {\scriptsize \textcolor{red}{action $3$}};
\end{tikzpicture}
} 

%% file: learning-short.tex
\paragraph{Online Learning with Bandit Feedback.}
There is a single principal interacting repeatedly with a single agent, whose type is drawn from an unknown type distribution $\mathcal{D}$. The interaction takes place over $\numrounds$ rounds.
In each round $\round$, the agent's type $\theta^\round$ is drawn from the underlying type distribution $\mathcal{D}$, independently.
Over all rounds, the principal has fixed rewards $\{\rew_j\}_{j \in [m]}$. 
The agent's type $\theta^\round$ determines the cost $c^\theta_i \geq 0$ of each action $i \in [n]$, as well as the probability distribution $\Prob^\theta_i$ over outcomes $j \in [m]$. It is assumed that both rewards and costs are bounded in $[0,1]$.%
\footnote{Since regret is an additive metric, we need to specify the range of the key quantities involved. Normalization to $[0,1]$ can be achieved through appropriate scaling, but also scales the regret with respect to the original unscaled instance.}

In each round $\round$, 
the principal posts a contract $\con^\round = (\pay^\round_1, \ldots, \pay^\round_m)$ (a non-negative payment for each outcome). The choice of contract may depend on what the principal has observed so far, and 
may be randomized. We consider two classes of contracts. In a (general) \emph{bounded} contract we have $\con^\round \in [0,1]^m$,
while a \emph{linear} contract is defined as before. 
After the principal has posted contract $\con^\round$, a type $\theta^\round$ is drawn from $\mathcal{D}$, the agent takes a best response action $i^{\star}(\theta^\round,\con^\round)$, and an outcome $j^\round$ is sampled from $\Prob_{i^{\star}(\theta^\round,\con^\round)}$. The principal learns about the outcome $j^\round$, receives the corresponding reward $r_{j^\round}$, and pays the agent the amount specified by contract $\con^\round$ for outcome $j^\round$.

The principal's goal is to minimize \emph{regret} with respect to the best single contract in hindsight.
To formally define this, let $\mathcal{T}$ denote a class of contracts. Let $U_P(\theta \mid \con)$ 
denote the expected principal utility for contract $\con$ when the agent's type is $\theta$, and let $\pi$ be a policy which maps each history $\mathcal{H}^{\round-1}$ to a distribution over contracts. Then 
\[
\mathsf{regret}(\pi,\mathcal{T}) := \sup_{\bar{\con} \in \mathcal{T}} \sum_{\round=1}^{\numrounds} \mathbb{E}_{\con^\round \sim \pi(\mathcal{H}^{\round-1})} \left(\mathbb{E}_{\theta^\round}[U_P(\theta^\round \mid \bar{\con})] - 
\mathbb{E}_{\theta^\round}[U_P(\theta^\round \mid \con^\round)] \right).
\]

\paragraph{Separation: General vs.~Linear Contracts.}
The main results of \cite{ZhuEtAl22} are near-tight bounds 
on the regret, showing that while the regret for (general) bounded contracts has to be essentially linear, linear contracts admit sublinear regret bounds.

\begin{theorem}[(\citet*{ZhuEtAl22})]\label{thm:zhu1}
~\\
(1)~
For (general) 
bounded contracts there is an online learning algorithm that incurs a regret of at most $\tilde{O}(\sqrt{m} \cdot \numrounds^{1-1/(2m+1)})$, and no online learning algorithm can incur a regret better than $\Omega(\numrounds^{1-1/(m+2)})$.

\noindent(2)~
For 
linear contracts there is an online learning algorithm that incurs a regret of at most $\tilde{O}(\numrounds^{2/3})$, and no online learning algorithm can incur a regret better than $\Omega(\numrounds^{2/3})$.
\end{theorem}

The intuitive reason for the sharp separation between (general) bounded contracts and linear contracts is the very different geometry of the principal’s utility in the two cases. For linear contracts (Figure~\ref{fig:upper-envelope-principal}), the optimal principal utility is a well-behaved, right-continuous function. In particular, there is always a “safe” direction: if $\alpha^\star$ is the optimal linear contract and we over-shoot by $\epsilon$, then we only lose $\epsilon$ utility. By contrast, for general (bounded) contracts, the principal’s utility admits only a weak continuity property: for each contract 
$\con$, there exists a direction (a cone) along which the utility does not drop too much. While this enables a cover-by-cones argument, yielding the exponential upper bound, the lower-bound construction shows there is no additional structure that would enable more efficient learning.

Notably, the impossibility for (general) bounded contracts already holds for a fixed agent (i.e, when $\mathcal{D}$ is a point-mass distribution), 
and requires exponentially many (in $m$) actions. The case of a fixed agent with a small action space is handled in \cite{BacchiocchiC0024}. An appealing open problem is whether the exponential lower bound can be circumvented under additional structure, e.g., MLRP 
(it can, for instances satisfying FOSD and CDFP 
\citep{ChenCDH24}).

%% file: bib-short.bib
@article{DuettingFT26,
    author       = {Paul D{\"{u}}tting and
                  Michal Feldman and
                  Inbal Talgam{-}Cohen},
    title        = {An Introduction to Contract Theory for Computer Scientists},
    journal      = {ACM SIGecom Exchanges},
    volume       = {24},
    number       = {1},
    pages        = {31--46},
    year         = {2026},
}

@article{HermalinK91,
    author = {Benjamin E. Hermalin and Michael L. Katz},
    title = {Moral Hazard and Verifiability: {T}he Effects of Renegotiation in Agency}, journal = {Econometrica},
    year = {1991}, 
    volume = {59},
    issue = {6}, 
    pages = {1735–1753}
}

@inproceedings{EzraFS24,
    author = {Tomer Ezra and Michal Feldman and Maya Schlesinger},
    title = {On the (in)approximability of Combinatorial Contracts},
    year = {2024},
    booktitle = {Proc.~of ITCS 2024},
    pages     = {44:1--44:22},
    publisher = {Schloss Dagstuhl--Leibniz-Zentrum für Informatik},
    address = {Wadern, Germany}
}

@inproceedings{DuettingFG23,
    title = {Combinatorial Contracts beyond Gross Substitutes},
    author = {Paul D\"utting and Michal Feldman and Yoav Gal-Tzur},
    booktitle = {Proc.~of SODA 2024},
    pages = {92--108},
    year = {2024},
    publisher = {SIAM},
    address = {Philadelphia, PA, USA}
}

@inproceedings{DuettingEFK24,
    author       = {Paul D\"utting and
    Tomer Ezra and
    Michal Feldman and
    Thomas Kesselheim},
    title        = {Multi-Agent Combinatorial Contracts },
    booktitle = {Proc.~of  SODA 2025},
    pages = {1857--1891},
    year = {2025},
    publisher = {SIAM},
    address = {Philadelphia, PA, USA}
}

@article{Feldman25,
  author       = {Michal Feldman},
  title        = {Combinatorial Contract Design: Recent Progress and Emerging Frontiers},
  journal      = {arXiv preprint arXiv:2510.15065},
  year         = {2025},
}

@inproceedings{sample-complexity,
  title = {The Pseudo-Dimension of Contracts},
  author = {Paul D\"utting and Michal Feldman and Tomasz Ponitka and Ermis Soumalias},
  booktitle = {Proc.~EC 2025},
  year = {2025},
  pages = {514--539},
  publisher = {ACM},
  address = {New York, NY, USA}
}

@article{Holmstrom82,
    author = {Bengt Holmstr\"om},
    title = {Moral hazard in teams},
    journal = {Bell J. Econ.}, 
    volume = {13},
    pages = {324–340}, 
    year = {1982}
}

@article{DuettingEFK23,
    author    = {Paul D\"utting and
    Tomer Ezra and
    Michal Feldman and
    Thomas Kesselheim},
    title     = {Multi-Agent Contracts},
    journal = {J. ACM},
    volume = {73},
    number = {2},
    pages = {13:1-13:37},
    year = {2026},
    note = {An earlier version appered in STOC 2023.}
}

@article{DuettingEFK21,
    author    = {Paul D\"utting and
    Tomer Ezra and
    Michal Feldman and
    Thomas Kesselheim},
    title     = {Combinatorial Contracts},
    journal = {SIAM J. Comput.},
    volume ={54},
    number = {6},
    pages = {1427-1455},
    year = {2025},
    note = {An earlier version appeared in FOCS 2021},
}

@article{AlonLST23,
    author       = {Tal Alon and
    Ron Lavi and
    Elisheva S. Shamash and
    Inbal Talgam{-}Cohen},
    title        = {Incomplete Information {VCG} Contracts for Common Agency},
    howpublished = {Technical Note},
    journal      = {Oper. Res.},
    volume = {72},
    number = {1},
    pages = {288--299}, 
    year = {2024},
    note = {An earlier version appeared in EC 2021.}
}

@inproceedings{GuruganeshSW21,
    title={Contracts under moral hazard and adverse selection}, 
    author={Guru Guruganesh and Jon Schneider and Joshua Wang},
    year={2021},
    booktitle={Proc.~of EC 2021},
    pages={563--582},
    publisher = {ACM},
    address = {New York, NY, USA}
}

@inproceedings{DuttingRT19,
    author    = {Paul D\"utting and
    Tim Roughgarden and
    Inbal Talgam{-}Cohen},
    title     = {Simple versus optimal contracts},
    booktitle = {Proc.~of EC 2019},
    pages     = {369--387},
    year      = {2019},
    note     = {Full version available at \url{https://arxiv.org/pdf/1808.03713}},
    publisher = {ACM},
    address = {New York, NY, USA}
}

@article{DuttingRT21,
    author    = {Paul D\"utting and
    Tim Roughgarden and
    Inbal Talgam{-}Cohen},
    title     = {The complexity of contracts},
    journal      = {{SIAM} J. Comput.},
    volume       = {50},
    number       = {1},
    pages        = {211--254},
    year         = {2021},
    note = {An earlier version appeared in SODA 2020.}
}

@article{GrossmanH83,
    author  = {Sanford J. Grossman and Oliver D. Hart},
    title   = {An analysis of the principal-agent problem},
    journal = {Econometrica},
    volume  = {51},
    number  = {1},
    year    = {1983},
    pages   = {7--45}
}

@inproceedings{VuongDPP23,
    author       = {Deo{-}Campo~Vuong, Ramiro  and
    Dughmi, Shaddin  and
    Patel, Neel  and
    Prasad, Aditya},
    title        = {On Supermodular Contracts and Dense Subgraphs},
    booktitle = {Proc.~of SODA 2024},
    pages = {109--132},
    year = {2024},
    publisher = {SIAM},
    address = {Philadelphia, PA, USA}
}

@inproceedings{EzraLR24,
    author       = {Tomer Ezra and
    Stefano Leonardi and
    Matteo Russo},
    title        = {Contract Design Beyond Hidden Actions},
    year         = {2026},
    booktitle    = {Proc.~of SODA 2026},
    publisher = {SIAM},
    pages = {2113-2133},
    address = {Philadelphia, PA, USA}
}

@inproceedings{KleinbergK18,
    author       = {Jon M. Kleinberg and
    Robert Kleinberg},
    title        = {Delegated Search Approximates Efficient Search},
    booktitle    = {Proc.~of EC 2018},
    pages        = {287--302},
    year         = {2018},
    publisher = {ACM},
    address = {New York, NY, USA}
}

@inproceedings{HartlineSLW22,
    author       = {Jason D. Hartline and
    Liren Shan and
    Yingkai Li and
    Yifan Wu},
    title        = {Optimal Scoring Rules for Multi-dimensional Effort},
    booktitle    = {Proc.~of COLT 2023},
    year         = {2023},
    pages        = {2624--2650},
    publisher = {PMLR}
}

@article{HoSV16,
    author    = {Chien{-}Ju Ho and
    Aleksandrs Slivkins and
    Jennifer Wortman Vaughan},
    title     = {Adaptive contract design for crowdsourcing markets: Bandit algorithms for repeated principal-agent problems},
    journal   = {J. Artif. Intell. Res.},
    volume    = {55},
    pages     = {317--359},
    year      = {2016},
    note = {An earlier version appeared in EC 2014.}
}

@inproceedings{KleinbergR19,
    author = {Kleinberg, Jon and Raghavan, Manish},
    title = {How do classifiers induce agents to invest effort strategically?},
    year = {2019},
    booktitle = {Proc.~of EC 2019},
    pages = {825--844},
    publisher = {ACM},
    address = {New York, NY, USA}
}

@article{Carroll15,
    title={Robustness and linear contracts},
    author={Carroll, Gabriel},
    journal={Am. Econ. Rev.},
    volume={105},
    number={2},
    pages={536--63},
    year={2015}
}

@inproceedings{BacchiocchiC0024,
    author       = {Francesco Bacchiocchi and
    Matteo Castiglioni and
    Alberto Marchesi and
    Nicola Gatti},
    title        = {Learning Optimal Contracts: How to Exploit Small Action Spaces},
    booktitle    = {Proc.~of ICLR 2024},
    year         = {2024},
    publisher = {PMLR}
}

@inproceedings{DuettingFGR24,
    title={When Contracts Get Complex: Information-Theoretic Barriers},
    author={Paul D\"utting and Michal Feldman and Yoav Gal-Tzur and Aviad Rubinstein},
    year={2026},
    booktitle={Proc.~of SODA 2026},
    pages = {6464-6493},
    publisher = {SIAM},
    address = {Philadelphia, PA, USA}
}

@inproceedings{AlonDLT23,
    title = {Bayesian Analysis of Linear Contracts},
    author = {Tal Alon and Paul D\"utting and Yingkai Li and Inbal Talgam-Cohen},
    booktitle = {Proc.~of EC 2023},
    year = {2023},
    page = {66},
    note = {Full version available at: \url{https://arxiv.org/abs/2211.06850}},
    publisher = {ACM},
    address = {New York, NY, USA}
}

@inproceedings{ezra2024sequentialcontracts,
    title={Contract Design for Sequential Actions}, 
    author={Tomer Ezra and Michal Feldman and Maya Schlesinger},
    year={2026},
    booktitle = {Proc.~of SODA 2026},
    publisher = {SIAM},
    pages = {6537--6570},
    address = {Philadelphia, PA, USA}
}

@article{CastiglioniM022,
    author    = {Matteo Castiglioni and
    Alberto Marchesi and
    Nicola Gatti},
    title     = {Designing Menus of Contracts Efficiently: The Power of Randomization},
    journal   = {Artif. Intell.},
    volume = {318},
    year = {2023},
    pages = {103881},
    note = {An earlier version appeared in EC 2022.}
}

@article{BabaioffFNW12,
    author = {Moshe Babaioff and Michal Feldman and Noam Nisan and Eyal Winter},
    title = {Combinatorial agency},
    journal = {J. Econ. Theory},
    volume = {147},
    number = {3},
    pages = {999--1034},
    year = {2012},
    note = {An earlier version appeared in EC 2006.}
}

@article{DuttingFT24,
  author       = {Paul D{\"{u}}tting and
                  Michal Feldman and
                  Inbal Talgam{-}Cohen},
  title        = {Algorithmic Contract Theory: {A} Survey},
  journal      = {Found. Trends Theor. Comput. Sci.},
  volume       = {16},
  number       = {3-4},
  pages        = {211--412},
  year         = {2024},
  note         = {\url{https://arxiv.org/abs/2412.16384}}
}

@article{NisanR01,
    title = {Algorithmic Mechanism Design},
    journal = {Games Econ. Behav.},
    author = {Noam Nisan and Amir Ronen},
    volume = {35},
    number = {1},
    pages = {166-196},
    year = {2001},
    note = {An earlier version appeared in STOC 1999.}
}

@article{Holmstrom79,
    author = {Bengt Holmstr\"om},
    journal = {Bell J. Econ.},
    title = {Moral Hazard and Observability},
    volume = {10}, 
    issue = {1}, 
    pages = {74-91},
    year = {1979}
}

@inproceedings{ZhuEtAl22,
    author       = {Banghua Zhu and
    Stephen Bates and
    Zhuoran Yang and
    Yixin Wang and
    Jiantao Jiao and
    Michael I. Jordan},
    title        = {The Sample Complexity of Online Contract Design},
    year         = {2023},
    booktitle = {Proc.~of EC 2023},
    pages        = {1188},
    publisher = {ACM},
    address = {New York, NY, USA}
}

@inproceedings{DuettingGSW23, 
    author = {Paul D\"utting and Guru Guruganesh and Jon Schneider and Joshua Wang},
    title = {Optimal No-Regret Learning of One-Sided {L}ipschitz Functions},
    booktitle = {Proc.~of ICML 2023},
    year = {2023},
    publisher = {PMLR}
}

@inproceedings{BabichenkoEtAl2024,
    author = {Yakov Babichenko and Inbal Talgam-Cohen and Haifeng Xu and Konstantin Zabarnyi},
    title = {Information design
in the principal-agent problem},
    booktitle = {Proc.~EC 2024},
    year = {2024},
    pages = {669-670},
    publisher = {ACM},
    address = {New York, NY, USA}
}

@inproceedings{AlonDPTT20,
    author    = {Tal Alon and
    Magdalen Dobson and
    Ariel D. Procaccia and
    Inbal Talgam{-}Cohen and
    Jamie Tucker{-}Foltz},
    title     = {Multiagent Evaluation Mechanisms},
    booktitle = {Proc.~of AAAI 2020},
    pages     = {1774--1781},
    year = {2020},
    publisher = {AAAI Press},
    address = {Washington, DC, USA}
}

@inproceedings{AlonDT21,
    author    = {Tal Alon and
    Paul D\"utting and
    Inbal Talgam{-}Cohen},
    title     = {Contracts with Private Cost per Unit-of-Effort},
    booktitle = {Proc.~of EC 2021},
    pages     = {52--69},
    year      = {2021},
    publisher = {ACM},
    address = {New York, NY, USA}
}

@article{DuettingFP23,
    author       = {Paul D\"utting and
    Michal Feldman and Daniel Peretz and Larry Samuelson
    },
    title        = {Ambiguous Contracts},
    journal      = {Econometrica},
    volume ={92}, 
    number = {6}, 
    pages = {1967–1992},
    note = {An earlier version appeared in EC 2023.},
    year = {2024}
}

@misc{DuettingFR25,
  author       = {Paul D\"utting and
                  Michal Feldman and
                  Yarden Rashti},
  title        = {Succinct Ambiguous Contracts},
  journal      = {CoRR},
  volume       = {abs/2503.02592},
  year         = {2025},
  eprinttype    = {arXiv},
  note = {Available at \url{https://arxiv.org/abs/2503.02592}}
}

@inproceedings{WangEtAl2023,
    author       = {Tonghan Wang and
    Paul D\"utting and
    Dmitry Ivanov and
    Inbal Talgam{-}Cohen and
    David C. Parkes},
    title        = {Deep Contract Design via Discontinuous Networks},
    year         = {2023},
    booktitle    = {Proc.~of NeuRIPS 2023},
    publisher = {PMLR}
}

@inproceedings{SaigET24,
    author       = {Eden Saig and
    Ohad Einav and
    Inbal Talgam{-}Cohen},
    title        = {Incentivizing Quality Text Generation via Statistical Contracts},
    booktitle    = {Proc.~of NeurIPS 2024},
    year         = {2024},
    publisher = {PMLR}
}

@inproceedings{ChenCDH24,
    author       = {Yurong Chen and
    Zhaohua Chen and
    Xiaotie Deng and
    Zhiyi Huang},
    title        = {Are Bounded Contracts Learnable and Approximately Optimal?},
    booktitle    = {Proc.~of {EC} 2024},
    year         = {2024},
    publisher = {ACM},
    address = {New York, NY, USA}
}

@inproceedings{CastiglioniCL+25,
    author       = {Matteo Castiglioni and
    Junjie Chen and
    Minming Li and
    Haifeng Xu and
    Song Zuo},
    title        = {A Reduction from Multi-Parameter to Single-Parameter {B}ayesian Contract Design},
    booktitle    = {Proc.~of SODA 2025},
    pages         = {1795--1836},
    year         = {2025},
    publisher = {SIAM},
    address = {Philadelphia, PA, USA}
}

@inproceedings{CastiglioniC25,
    author       = {Matteo Castiglioni and Junjie Chen},
    title        = {Hiring for An Uncertain Task: Joint Design of Information and Contracts},
    booktitle = {Proc.~of SODA 2025},
    year = {2025},
    pages = {1758-1794},
    publisher = {SIAM},
    address = {Philadelphia, PA, USA}
}
